# Single crystal growth and characterizations of $Cu_{0.03}TaS_2$ superconductors

X.D. Zhu[a], Y.P. Sun[a,b,]*, X.B. Zhu[a], X. Luo[a], B.S. Wang[a], G. Li[a], Z.R. Yang[a], W.H. Song[a], J. M. Dai[a]

[a]Key Laboratory of Materials Physics, Institute of Solid State Physics, Chinese Academy of Sciences, Hefei 230031, People's Republic of China

[b]High Magnetic Field Laboratory, Chinese Academy of Sciences, Hefei 230031, People's Republic of China

**Abstract**

Single crystal of $Cu_{0.03}TaS_2$ with low copper intercalated content was successfully grown via chemical vapor transport method. The structural characterization results show that the copper intercalated $2H$-$Cu_{0.03}TaS_2$ single crystal has the same structure of the $CdI_2$-type structure as the parent $2H$-$TaS_2$ crystal. Electrical resistivity and magnetization measurements reveal that $2H$-$Cu_{0.03}TaS_2$ becomes a superconductor below 4.2 K. Besides, electrical resistivity and Hall effects results show that a charge density wave transition occurs at $T_{CDW}$ = 50 K.

* Corresponding author. Tel.: +86 551 559 2757; Fax: +86 551 559 1434.

*E-mail address:* ypsun@issp.ac.cn

# 1. Introduction

Diverse physical properties such as Charge-Density-Wave (CDW) and superconductivity in layered transition metal dichalcogenides (TMDC) have been abroad studied [1, 2]. TMDC have a general formula $TX_2$, where T is usually a transition metal atom from group of IVB, VB, VIB of the periodic table of elements and X is one of sulfur, selenium, or tellurium. The layered TMDC can be regarded as stacking two-dimensional X-T-X sandwiches. The bonding within each sandwich is covalent, while the bonding between them is weak van der Waals type. The crystal structures of the layered TMDC are usually described as $1T$, $2H$, $3R$, $4H_a$, $4H_b$, $6R$ phases [1, 2]. The integer denotes the number of X-T-X layers per unit cell perpendicular to the layers, while $T$, $H$, and $R$ denote trigonal, hexagonal, rhombohedral symmetries, respectively.

Introducing foreign atoms or molecules between the weak coupling sandwiches or layers is the process of intercalation. Intercalation not only increases the layer separation, but also provides a powerful way to tune the electronic structures of the host materials. Intercalated TMDC, graphite, and layered structural transition metal nitrides have received considerable attention due to their special structures and transport properties especially for superconductivities [2-4]. New interests have been aroused since the recent discoveries of superconductivities in $YbC_6$, $CaC_6$ and $Cu_xTiSe_2$ [5-7].

$2H$-$TaS_2$ is one of the layered TMDC [8]. Metallic $2H$-$TaS_2$ undergoes a CDW transition at 78K, and becomes a superconductor below 0.8 K [8]. Many inorganic atoms and organic molecules have been intercalated into $TaS_2$ [3, 9-14]. Intriguingly, superconductivity has been found in $Py_{1/2}TaS_2$ [9], $A_{0.33}TaS_2$ (A = Li, Na, K, Rb, Cs) [14], and $2H$-$Fe_xTaS_2$ ($x$ = 0.05) et al. [15]. Recently, the competition driven by intercalation between superconductivity and CDW in $Na_xTaS_2$ has been discovered [16]. Copper intercalations in $TaS_2$ have been achieved by chemical vapor transport (CVT) method to produce $Cu_{1/2}TaS_2$ [17], by direct reaction from copper and $TaS_2$ to produce single phase $Cu_xTaS_2$ (0.33 < $x$ < 0.75) [18], and by electro-chemical method [19]. However, superconductivity has not been observed in any high copper intercalation $Cu_xTaS_2$ [17-19]. Further, there is also no any report about lower copper intercalation in $Cu_xTaS_2$. Thus, it is interesting to explore whether low copper intercalation in $Cu_xTaS_2$ can induce superconductivity or not. The low copper intercalation experiments were carried out in order to search for possible superconductivity in $Cu_xTaS_2$ ($x < 0.33$). In this paper, we report the single crystal growth of $Cu_{0.03}TaS_2$ with low copper intercalated content via CVT method. Electrical resistivity and magnetization measurements reveal that $2H$-$Cu_{0.03}TaS_2$ becomes a superconductor below 4.2 K. Besides, resistivity and Hall effects results show that a charge density wave transition occurs at $T_{CDW}$ = 50 K.

## 2. Experimental procedure

Single crystal $Cu_{0.03}TaS_2$ was grown via CVT method. Firstly, polycrystalline $TaS_2$ powder was synthesized by solid state reaction. Stoichiometric amounts of Ta and S powders were mixed and sealed in an evacuated quartz tube. The tube was heated at 900°C for 4 days. Secondly, Cu, $TaS_2$ powders in mol ratio of 0.5:1, and 100 mg iodine were mixed and sealed in an evacuated quartz tube with a 12 mm diameter and a 19 cm length. The tube was inserted in a two-zone horizontal tube furnace with a source-zone temperature of 1000°C and a growth-zone temperature of 900°C for 10 days. Both zones were decreased to 440°C in several days, and then cooled to room temperature. Dark blue, mirror-like plates in a typical size of $3 \times 3 \times 0.2$ mm$^3$ were obtained, and the photograph is shown in Fig. 1. F. R. Gamble et al. has reported that $2H$-$TaS_2$ single crystals grown via CVT method were black and somewhat wrinkle [9].

The grown $Cu_{0.03}TaS_2$ plates were cleaned by ultrasonic in supersaturated aqueous solutions of KI, de-ionized water, and alcohol, respectively. The composition of $Cu_{0.03}TaS_2$ single crystal was characterized by energy dispersive x-ray spectroscopy (EDS) and inductively coupled plasma atomic emission spectrometry (ICP-AES, Atomscan Advantage). The crystal structure and phase purity were examined by powder and single-crystal X-ray diffraction pattern (XRD) using a Philips X' pert PRO x-ray diffractometer with Cu K$\alpha$ radiation at room temperature. The electrical resistivity was performed by the standard four-probe method using a Quantum Design Physical Property Measurement System (PPMS) (1.8 K $\leq$ T $\leq$ 400 K, 0 T $\leq$ H $\leq$ 9 T). The Hall effects experiments were also performed using PPMS. Magnetization measurements as a function of temperature were performed in a Quantum Design Superconducting Quantum Interference Device (SQUID) system (1.8 K $\leq$ T $\leq$ 400 K, 0 T $\leq$ H $\leq$ 5 T).

## 3. Results and discussions

*3.1. Chemical analysis*

The EDS pattern for the grown single crystal is shown in Fig. 2. The estimated mol ratio of Ta : S is about 1 : 2. Though the copper peak is evident in the EDS spectrum, the accurate copper content can not be obtained because the copper content is too small. Thus, the copper content is determined by ICP-AES. The determined mol ratio of the Cu : Ta is 0.03 : 1.

*3.2. Structural analysis*

The XRD patterns for $Cu_{0.03}TaS_2$ and $2H$-$TaS_2$ single crystals are shown in Fig. 3a. It can be observed that the orientations of the crystal surfaces are both (00l) planes. The magnification plots of (006) peaks are shown in the inset of Fig. 3a. Obviously, the peak positions of $Cu_{0.03}TaS_2$ shift to lower $2\theta$ value. According to the Bragg equation *nλ = 2dsinθ*, the copper intercalation leads to expansion of the lattice constant *c*.

Several single crystals were crushed into powders, which were used in the powder XRD experiment. The powder XRD patterns of crushed $Cu_{0.03}TaS_2$ and $2H$-$TaS_2$ crystals are shown in Fig. 3b. All the peaks can be well indexed to the $2H$ structure [20]. The present $Cu_{0.03}TaS_2$ is single phase with no detectable secondary phases. The lattice constants were obtained from powder XRD patterns. The lattice constant of $Cu_{0.03}TaS_2$ is almost equal to that of $2H$-$TaS_2$ ($a$ = 3.31Å), while the lattice constant *c* of $Cu_{0.03}TaS_2$ is 12.13(7) Å, which is slightly larger than that of $2H$-$TaS_2$ ($c$ = 12.08(0) Å).

*3.3. Co-existence of superconductivity and charge density wave*

The temperature dependence of electrical resistivity in *ab* plane ($\rho_{ab}$-*T*) for $2H$-$Cu_{0.03}TaS_2$ is plotted in Fig. 4. Obviously, there is a kink around *T* = 50 K, which can be attributed to CDW transition. The inset shows the enlarged view of the $\rho_{ab}$-*T* curve near the superconducting region. The resistivity sharply drops to zero around *T* = 4.2 K, which indicates the superconductivity. The onset superconducting transition temperature is 4.2 K, and the transition width (10%-90%) is 0.2 K.

The temperature dependence of dc magnetic susceptibility for $Cu_{0.03}TaS_2$ is plotted in Fig. 5. The magnetic field of 2 Oe was applied parallel to the *ab* plane of a $Cu_{0.03}TaS_2$ plate. A sharp drop of magnetization at 4.2 K was observed for both zero-field-cooling (ZFC) and field-cooling (FC) measurements, which further confirms the existence of superconductivity. The ZFC curve shows an almost perfect shielding effect, while the magnetic flux exclusion fraction estimated from the FC curve is as low as ~4% at 2 K, using $-4\pi\chi_v$ to estimate the volume fraction. Similar phenomena have been reported in other intercalated compounds such as $Py_{1/2}TaS_2$ [21], $YbC_6$ and $CaC_6$ [5, 6], $LiPd_3B_2$ [22]. Prober *et al.* suggested that the small FC magnetization was due to the complicated flux trapping effect in these intercalated compounds [23]. Thus, it is suggested that the bulk superconductivity is found in $Cu_{0.03}TaS_2$.

The temperature dependence of the Hall ($R_H$) coefficient for $Cu_{0.03}TaS_2$ is shown in the Fig. 6. The $R_H$ versus $T$ is determined from the Hall resistivity ($\rho_{xy}$) results using $R_H = \rho_{xy}/H$. The inset in Fig. 6 shows the magnetic field dependence of the $\rho_{xy}$ measured at different temperatures. The $R_H$ increases from ~ $2.8 \times 10^{-4} cm^3/C$ above 60 K to ~$5.5 \times 10^{-4} cm^3/C$ below 30K. In contrast, the sign of the $R_H$ of matrix $2H$-$TaS_2$ even change from positive to negative near the CDW transition, which was reported by A. H. Thompson et al [21]. The change of the $R_H$ of $Cu_{0.03}TaS_2$ in the region of 30 K< $T$ < 60 K confirms the CDW transition revealed by the resistivity results. Apparently, copper intercalation suppresses the $T_{CDW}$ from $T_{CDW}$ ~ 78 K ($2H$-$TaS_2$) to $T_{CDW}$ ~ 50 K. According to the formula $n_H = -\dfrac{1}{eR_H}$, the reduction of carrier density is consistent with the partial gapping of the Fermi surface due to the occurrence of CDW transition.

4. **Conclusion**

In summary, $Cu_{0.03}TaS_2$ single crystals with low copper intercalated content was successfully grown via CVT. Structure analysis shows that the low copper intercalated $Cu_{0.03}TaS_2$ and the parent compound $2H$-$TaS_2$ share the same structure. Electrical resistivity and magnetization measurements reveal that $2H$-$Cu_{0.03}TaS_2$ becomes a superconductor below 4.2 K. Besides, electrical resistivity and Hall effects results show that a charge density wave transition occurs at $T_{CDW}$ = 50 K.

**Acknowledgements**

This work was supported by the National Key Basic Research under contract No. 2006CB601005, 2007CB925002, and the National Nature Science Foundation of China under contract No.10774146, 10774147 and Director's Fund of Hefei Institutes of Physical Science, Chinese Academy of Sciences.

**References**

[1] J. A. Wilson and A. D. Yoffe, Adv. Phys. 18 (1969) 193.

[2] R. H. Friend and A. D. Yoffe, Adv. Phys. 36 (1987) 1.

[3] M. S. Dresselhaus and G. Dresselhaus, Adv. Phys. 56 (2002) 1.

[4] S. Yamanaka, Annu. Rev. Mater. Sci. 30 (2000) 53.

[5] T. E. Weller, T. E. Weller, M. Ellerby, S. S. Saxena, R. P. Smith, and N. T. Skipper, Nature Physics 1 (2005) 30.

[6] N.E mery, C. Hérold, M. d'Astuto, V. Garcia, Ch. Bellin, J. F. Marêché, P. Lagrange, and G. Loupias, Phys. Rev. Lett. 95 (2005) 087003.

[7] E. Morosan, H. W. Zandbergen, B. S. Dennis, J. W. G. Bos, Y. Onose, T. Klimczuk, A. P. R amirez, N. P. Ong and R. J. Cava, Nature Physics 2 (2006) 544.

[8] J. M. E. Harper, T. H. Geballe and F. J. Di Salvo, Phys. Rev. B 15 (1977), 2943.

[9] F. R. Gamble, F. J. Di Salvo, R. A. Klemm and T. H. Geballe, Science 168 (1970) 568.

[10] D. W. Murphy, F. J. Di Salvo, G. W. Hull, and J. V. Waszczak，Inorganic Chemistry 15 (1976) 17.

[11] R. M. Fleming and R. V. Coleman, Phys. Rev. Lett. 34 (1975) 1502.

[12] A. Schlicht, A. Lerf and W. Biberacher, Synthetic Metals 102 (1999) 1483.

[13] S. F. Meyer, R. E. Howard, G. R. Stewart, J. V. Acrivos, and T. H. Geballe, J. Chem. Phys. 62 (1975) 4411.

[14] A. Lerf, F. Sernetz, W. Biberacher, R. Schöllhorn, Mat. Res. Bull. 14 (1979) 797.

[15] R. M. Fleming and R. V. Coleman, Phys. Rev. Lett. 34 (1975) 1502.

[16]. L. Fang, Y. Wang, P. Y. Zou, L. Tang, Z. Xu, H. Chen, C. Dong, L. Shan, and H. H. Wen, Phys. Rev. B 72 (2005) 014534.

[17] R. de Ridder, G. van Tendeloo, J. van Landuyt, D. van. Dyck, and S. Amelinckx, Phys. Stat. Sol. (a) 37 (1976) 691.

[18] T. Uchida, S. Sato, M. Wakihara and M. Tanigychi, Nippon Kagaku Kaishi 10 (1991) 1306.

[19] C. Ramos, A. Lerf and T. Butz, Hyperfine Interactions 61 (1990) 1209.

[20] A. Meetsm, A. Meetsma, G. A. Wiegers, R. J. Haange and J. L. de Boer, Acta. Cryst. C 46 (1990) 1598.

[21] A. H. Thompson, F. R. Gamble and R. F. Koehler, Jr. Phys. Rev. B 5 (1972) 2811.

[22] K. Togano, P. Badica, Y. Nakamori, S. Orimo, H. Takeya, and K. Hirata, Phys. Rev. Lett. 93 (2004) 247004.

[23] D. E. Prober and M. R. Beasley and R. E. Schwall, Phys. Rev. B 15 (1977) 5245.

**Figure Captions:**

Fig. 1. Photograph of $Cu_{0.03}TaS_2$ single crystal.

Fig. 2. The EDS pattern for $Cu_{0.03}TaS_2$.

Fig. 3. The single crystal XRD patterns for $Cu_{0.03}TaS_2$ and $2H$-$TaS_2$; Inset: the magnification plot of single crystal XRD patterns; (b) The powder XRD patterns for $Cu_{0.03}TaS_2$ and $2H$-$TaS_2$.

Fig. 4. Temperature dependence of the in-plane resistivity ($\rho_{ab}$) for $Cu_{0.03}TaS_2$. The inset shows the $\rho_{ab}$-$T$ curve near the superconducting region. The arrows show the charge density wave transition temperature ($T_{CDW}$) and the onset transition temperature ($T_{Conset}$).

Fig. 5. The temperature dependence of ZFC and FC susceptibility $\chi_g$ measured in an applied field of 2 Oe parallel to *ab* plane for $Cu_{0.03}TaS_2$: FC (filled circles); ZFC (open circles).

Fig. 6. Temperature dependence of the Hall coefficients ($R_H$) for $Cu_{0.03}TaS_2$. The inset shows the measured Hall resistivity ($\rho_{xy}$) versus $H$ measured at different temperatures.

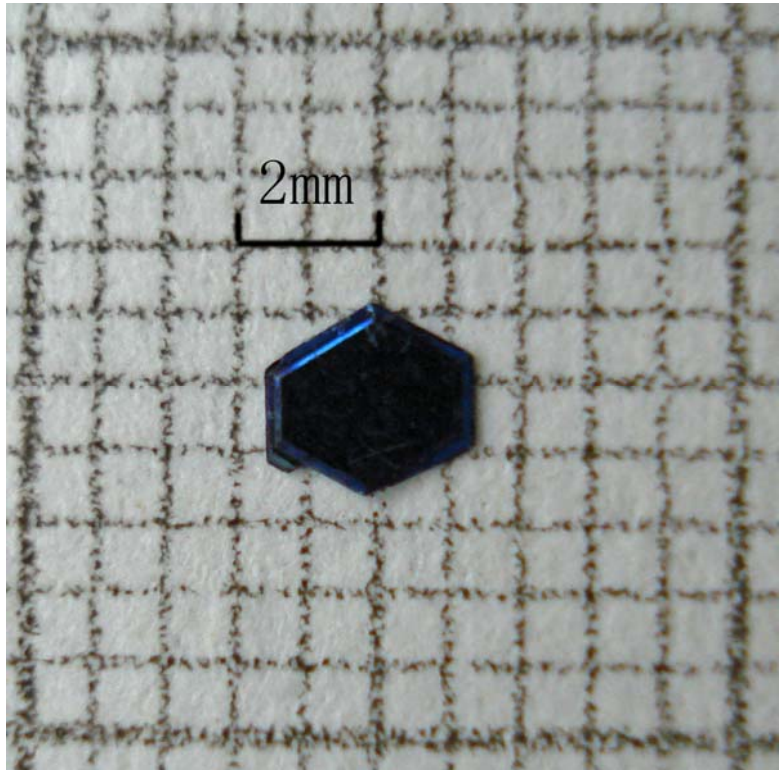

Fig. 1    X D.Zhu et al

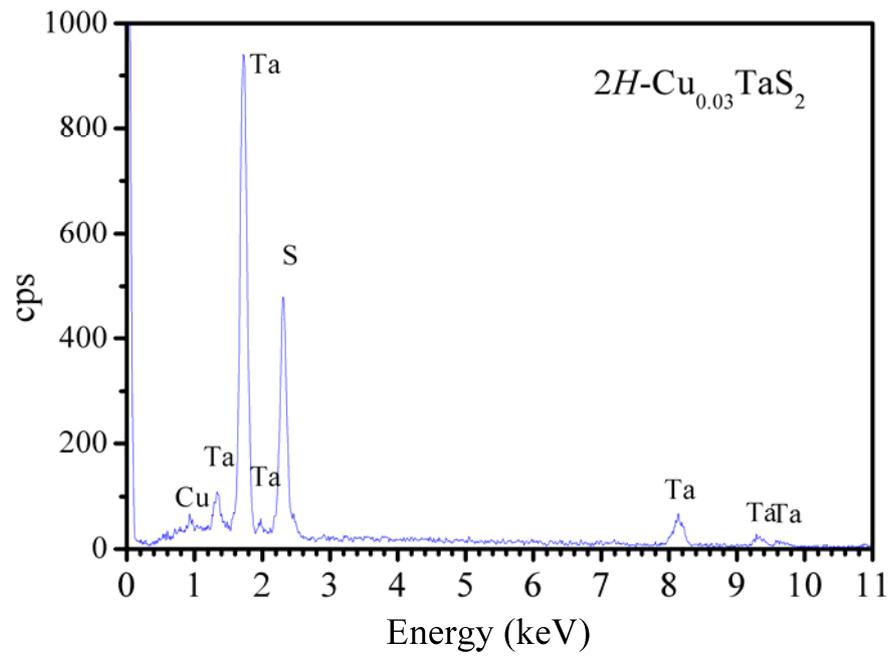

Fig. 2 X D.Zhu et al

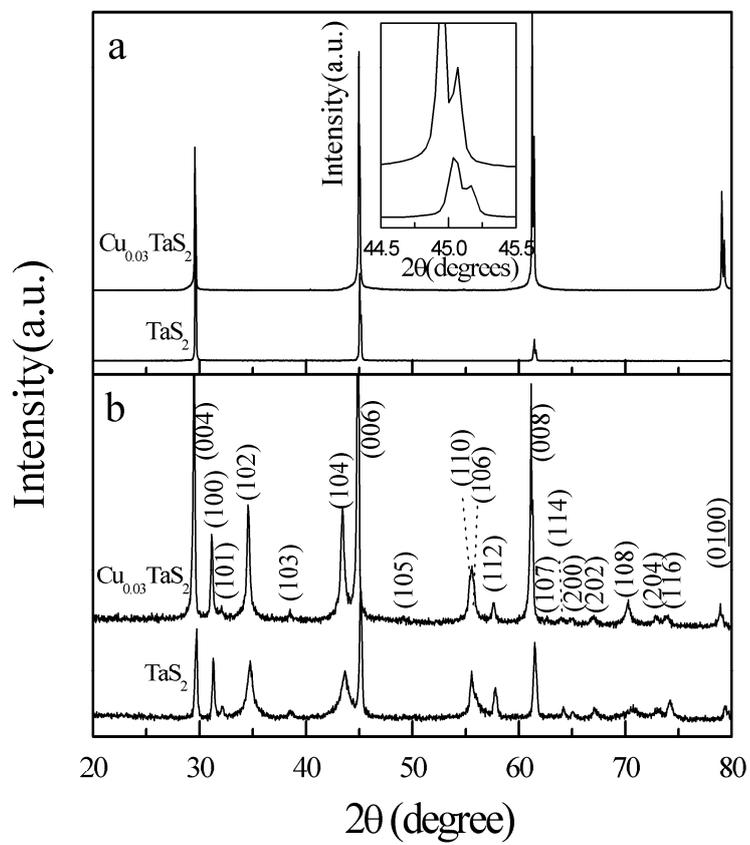

Fig. 3 X D.Zhu et al

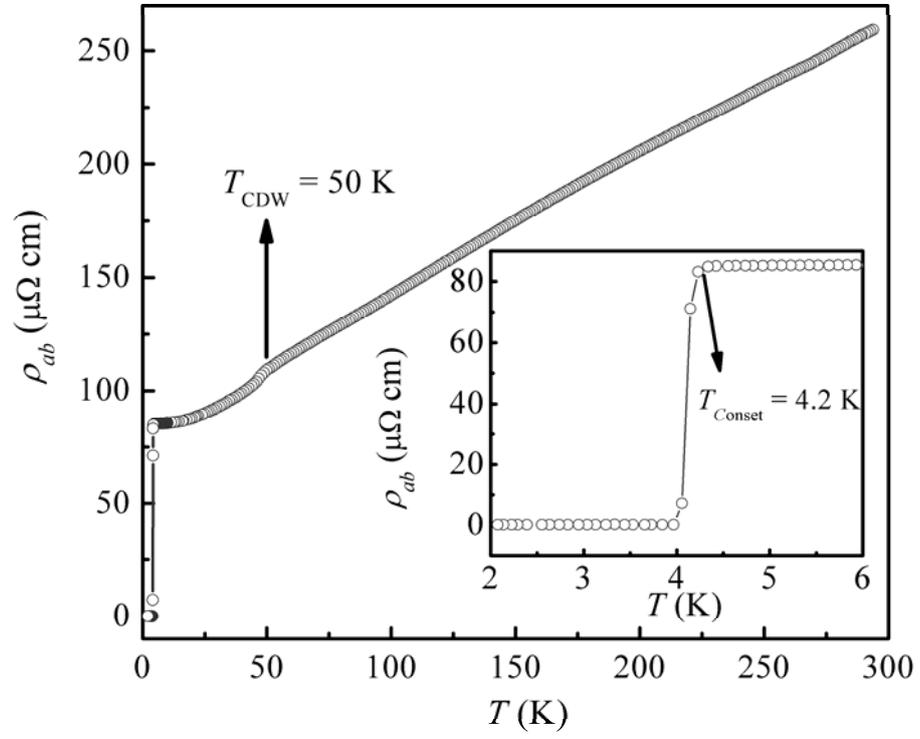

Fig. 4 X D.Zhu et al

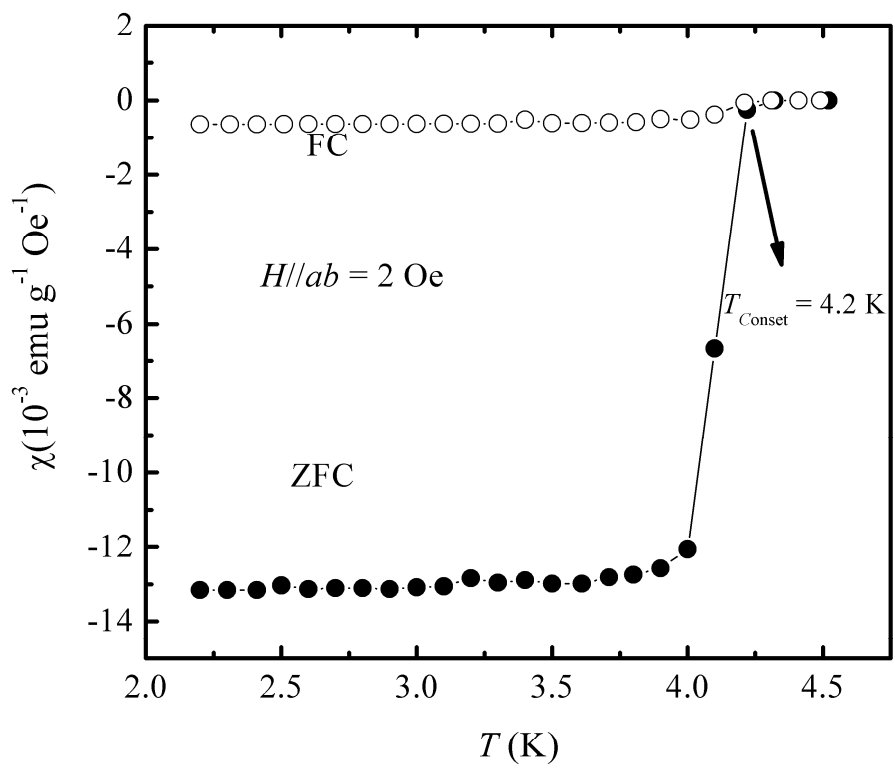

Fig. 5 X D.Zhu et al

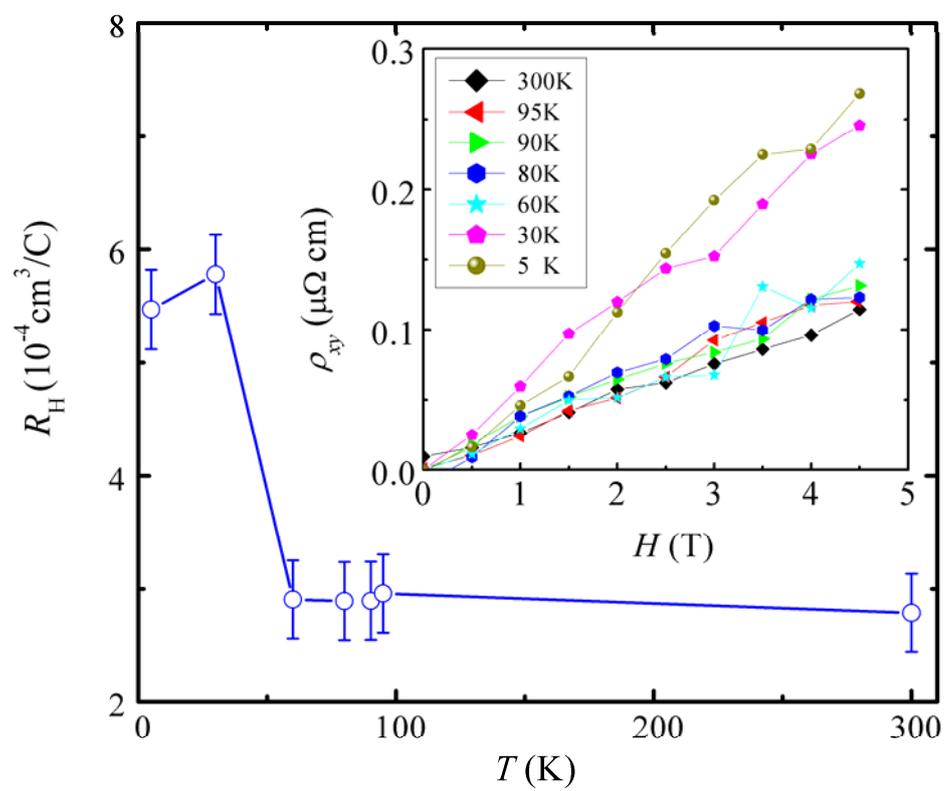

Fig. 6 X D.Zhu et al